# Femtosecond carrier dynamics and saturable absorption in graphene suspensions

Sunil Kumar,[1,2] M. Anija,[1,2] N. Kamaraju,[1,2] K. S. Vasu,[1] K. S. Subrahmanyam,[3]
A. K. Sood,[1-3,a)] and C. N. R. Rao[3]

[1]Department of Physics, Indian Institute of Science, Bangalore 560012, India
[2]Center for Ultrafast Laser Applications, Indian Institute of Science, Bangalore 560012, India
[3]Chemistry and Physics of Materials Unit and International Center for Materials Science,
Jawaharlal Nehru Center for Advanced Scientific Research, Jakkur P.O., Bangalore 560064, India

Nonlinear optical properties and carrier relaxation dynamics in graphene, suspended in three different solvents, are investigated using femtosecond (80 fs pulses) Z-scan and degenerate pump-probe spectroscopy at 790 nm. The results demonstrate saturable absorption property of graphene with a nonlinear absorption coefficient, $\beta$, of ~2 to $9 \times 10^{-8}$ cm/W. Two distinct time scales associated with the relaxation of photoexcited carriers, a fast one in the range of 130-330 fs (related to carrier-carrier scattering) followed by a slower one in 3.5-4.9 ps range (associated with carrier-phonon scattering) are observed.

Graphene is a two-dimensional carbon nanomaterial which has received tremendous interest in recent years owing to its various remarkable properties and applications in modern electronics and photonics [1,2]. Ultrafast degenerate and nondegenerate pump-probe measurements on single and multilayer epitaxial [3,4] or exfoliated [5] graphenes deposited on a substrate have shown two types of dynamics of the carriers: a fast component of the order of ~100 fs attributed to the intraband carrier-carrier scattering and a slower component ~2 ps associated with carrier-phonon scattering. In the degenerate pump-probe studies on single and multilayer graphene films grown on a SiC substrate using 85 fs laser pulses centered at 790 nm, a positive change in the transient differential transmission of the probe was observed with two relaxation times, the faster one in the range of 70-120 fs and a slower one between 0.4-1.7 ps [3]. Similar carrier relaxation dynamics was obtained in the nondegenerate pump-probe experiments on exfoliated graphene films on $SiO_2$/Si substrate [5]. On a few layer thick graphene film on SiC, nondegenerate pump-probe experiments [4] using 800 nm pump showed that the sign of the differential transmission signal is positive over the entire probe spectral range of 1.1 to 2.6 μm but becomes negative after 2 ps if the probe wavelength falls between 1.78 and 2.35 μm. The initial positive part of the signal within 150 fs has been described in terms of thermalization and emission of high energy-phonons followed by a slow decay of the order of a few ps determined by electron-acoustic phonon scattering. Nonlinear optical properties of graphene have been reported recently in the nanosecond (ns) and picosecond (ps) regimes [6,7]. Using 35 ps laser pulses centered at 532 nm, it has been shown that the nonlinear response of graphene oxide suspensions changes from saturable absorption at low intensity (2.1 $GW/cm^2$) to reverse saturable absorption or optical limiting at higher intensities (>4.5 $GW/cm^2$) [6]. In comparison, in the ns regime suspensions of graphene oxide and functionalized graphene in dimethylformamide showed optical limiting property at all values of intensities above 0.6 $GW/cm^2$ [6,7].

We have carried out femtosecond (80 fs) Z-scan and degenerate pump-probe experiments at 790 nm to study the nonlinear optical response and carrier dynamics in colloidal suspensions of graphene which have not been investigated hitherto. Graphene in solution phase has an advantage that the carrier scattering mechanisms are not influenced by the underlying substrate. From the Z-scan experiments at intensity ~16 $GW/cm^2$ we observe a saturable absorption behavior. The graphene samples studied by us show positive change in the differential transmission of the probe in the pump-probe experiment along with two component relaxation dynamics in the range of 130-330 fs and 3.5-4.9 ps associated with the carrier-carrier and carrier-phonon scattering processes in graphene.

Solution-phase and chemical exfoliation techniques provide a low-cost, high-yield method for mass production of graphene [8-14] as compared to commonly used methods such as micromechanical cleavage and epitaxial growth. For our study, graphene suspensions were prepared in three different solvents, distilled water, dimethylformamide (DMF) and tetrahydrofuran (THF). The results have been compared with those found with an aqueous suspension of graphene oxide. Water suspension of graphene oxide (GO) was first prepared from exfoliated graphite by using modified Hummers method [11]. Graphene oxide layers were further reduced by anhydrous hydrazine ($N_2H_4$) to obtain the graphene (reduced graphene oxide) suspension in water (G-$H_2O$). Raman spectra of the G-$H_2O$ showed a single 2D peak at 2686 $cm^{-1}$ characteristic of single layer graphene. The concentration of the GO and G-$H_2O$ suspensions were ~80 μg/ml and 60 μg/ml, respectively. Suspensions of functionalized graphene in DMF and THF were obtained and characterized as described in the literature [13,14]. In the later samples, the suspensions contained few-layer graphene flakes. The concentration of graphene in DMF (G-DMF) and in THF (G-THF) was ~250 μg/ml. 100 μl of each sample was taken in a 1 mm optical path length quartz cuvette to be used in the pump-probe and Z-scan experiments. The measured values of the linear absorption coefficient of the suspensions are given in Table 2. We have used a Ti:sapphire regenerative femtosecond amplifier system (50 fs, 790 nm, 1 kHz, Spitfire, Spectra physics). The pulse-width in our experiments was ~80 fs at the sample point measured by cross-correlation in a thin BBO crystal. In the pump-probe experiments, the pump pulse intensity was fixed at ~2.0 GW/cm$^2$ and probe was fixed at ~0.17 GW/cm$^2$. The polarizations of the two pulses were kept orthogonal to each other and a polarizer crossed with the pump polarization was placed before the detector to avoid any pump scattering. The pump beam was modulated at ~383 Hz with the help of a chopper and the change in probe transmitted intensity was measured with a lock-in amplifier. In the Z-scan experiments, a focusing lens was used to achieve a continuous variation of intensity across the focal point (diameter 50 μm) changing from a value as low as ~4.0x10$^5$ W/cm$^2$ (far away from the focal point) to a value as high as ~1.6x10$^{10}$ W/cm$^2$ (at the focal point).

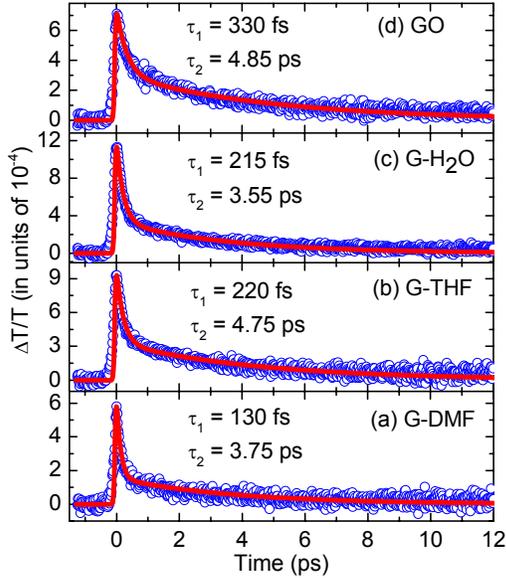

FIG. 1. (Color online) Transient differential transmission spectra for the graphene suspensions as a function of probe delay in the degenerate pump-probe experiment at 790 nm (1.57 eV): (a) G-DMF, (b) G-THF, (c) G-H$_2$O and (d) GO. The solid lines are fits to bi-exponentially decaying functions convoluted with the cross correlation of the pump and probe pulses.

TABLE I. Parameters obtained from the transient differential transmission spectra of the graphene suspensions.

| Sample | $S$ | $\beta$ (cm/W) | $\tau_1$ (fs) | $\tau_2$ (ps) |
|---|---|---|---|---|
| G-DMF | 5.8x10$^{-4}$ | 1.2x10$^{-8}$ | 130 | 3.75 |
| G-THF | 9.3x10$^{-4}$ | 1.9x10$^{-8}$ | 220 | 4.75 |
| G-H$_2$O | 11.3x10$^{-4}$ | 9.4x10$^{-8}$ | 215 | 3.55 |
| GO | 7.1x10$^{-4}$ | 4.4x10$^{-8}$ | 330 | 4.85 |

Transient differential transmission spectra ($\Delta T(t)/T$) of the four suspensions G-DMF, G-THF, G-H$_2$O and GO are presented in Fig. 1. Here $\Delta T(t)$ is the time dependent change in probe transmission, induced by the pump at time $t$ after the pump excitation and $T$ is the probe transmission in the absence of the pump. At $t = 0$, i.e. immediately after the pump excitation, the transmission rapidly increases and then recovers in two distinct time scales, the fast time constant arising from the intraband carrier-carrier scattering and the slow one in terms of carrier-phonon scattering processes [3-5,15]. Our data are fitted (solid lines in Fig. 1) using a bi-exponentially decaying function, $\Delta T(t)/T = A_1\exp(-t/\tau_1) + A_2\exp(-t/\tau_2)$, convoluted with the cross-correlation of the pump and probe pulses. The parameters are given in Table 1. The faster time constant ($\tau_1$) is as small as 130 fs for G-DMF and the slower one ($\tau_2$) is as large as 4.85 ps for GO. The major contribution (>60%) to the recovery of the differential transmission signal is due to the fast component. The time constants for GO (Fig. 1d) are ~ 330 fs and 4.85 ps which should be compared with ~215 fs and 3.55 ps for G-H$_2$O (Fig. 1c). A larger value of $\tau_1$ associated with the intraband carrier-carrier scattering in GO as compared to that in G-H$_2$O can arise because the carrier density in GO is much smaller than in G-H$_2$O as reflected in the ratio of the electrical resistance of the two samples ($R_{GO}/R_{G-H2O} \sim 10^3$). Similar reasoning will justify the smaller $\tau_2$ associated with the carrier-phonon scattering processes [4] for G-H$_2$O. We note that for our graphene suspensions, $\tau_2$ is in the range of 3.5 to 4.9 ps, much larger as compared to the graphene layers on a substrate: ~1.4 ps for 6 layered graphene on SiC [3] and ~2.5 ps for single layer of graphene on SiO$_2$/Si [5]. This difference can arise due to the absence of scattering of the carriers at the graphene-substrate interface.

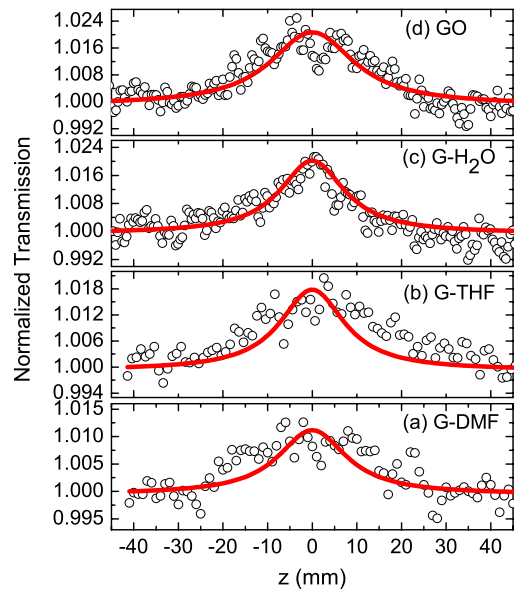

FIG. 2. Open aperture Z-scan data of the graphene suspensions: (a) G-DMF, (b) G-THF, (c) G-H$_2$O and (d) GO. The solid lines are the theoretical fits.

The positive sign of ΔT/T at $t = 0$ (denoted by $S$) is generally referred to as bleaching of the ground state of the system. The related phenomenon is observed as saturable absorption in the open aperture (OA) Z-scan experiments, as discussed later. The magnitude of the nonlinear absorption coefficient, $\beta$ can be approximately determined from $S$ by using the relation [16] $\beta = \dfrac{\ln(1+S)}{L_{eff} I_{pump}}$, where $L_{eff}$ = $L$(cuvette optical length)x$C$(graphene concentration) is the effective optical path length of the sample and $I_{pump}$ is the pump intensity. The values of $\beta$ are given in Table 1 for all the four suspensions. Note that Im$\chi^{(3)}$ (imaginary part of the third order nonlinear optical susceptibility $\chi^{(3)}$) can be easily calculated from $\beta$ using the relation $\operatorname{Im}\chi^{(3)} = \left(\dfrac{10^{-7} c \lambda n^2}{96\pi^2}\right)\beta$ where Im$\chi^{(3)}$ is in esu, $\beta$ is in cm/W; $c$ is speed of light in vacuum, $\lambda$ is the laser wavelength and $n$ is the linear refractive index of the sample ($n$ is taken as 1.5).

Open aperture (OA) Z-scan data for the four graphene suspensions are shown in Fig. 2. The nonlinear optical constant, $\beta$ and the saturation intensity, $I_s$ are deduced after fitting the data with the theory given in ref. [17]. The parameters $\beta$ and $I_s$ obtained from the fitting are given in Table 2. The values of $\beta$ are consistent with those evaluated from the pump-probe data shown in Table 1. A figure of merit for the third order optical nonlinearity is defined as FOM $= |\operatorname{Im}\chi^{(3)}|/\alpha$ where $\alpha$ is the linear absorption coefficient. The measured values of FOM are given in Table 2. For all the four graphene suspensions the values of the FOM are comparable with those of single walled carbon nanotube (SWNT) suspensions [17-18] and double walled carbon nanotube (DWNT) suspensions [19]. Under the same experimental conditions as used for OA, no appreciable nonlinear refraction was observed in the closed aperture (CA) Z-scan, similar to the results using ps and ns laser pulses [6]. We have checked that the pump-probe and Z-scan experiments carried out on the pure solvents (water, DMF and THF) did not show any response under the same experimental conditions.

TABLE II. Optical constants of the graphene suspension measured from OA Z-scan experiments. $\alpha$ is the linear absorption coefficient measured separately.

| Sample | $\alpha$ (cm$^{-1}$) | $\beta$ (cm/W) | $I_s$ (W/cm$^2$) | FOM (esu.cm) |
|---|---|---|---|---|
| G-DMF | 2.2x10$^3$ | 2.0x10$^{-8}$ | 2.3x10$^{10}$ | 5.0x10$^{-15}$ |
| G-THF | 3.4x10$^3$ | 2.0x10$^{-8}$ | 2.6x10$^{10}$ | 3.3x10$^{-15}$ |
| G-H$_2$O | 1.4x10$^4$ | 9.0x10$^{-8}$ | 2.5x10$^{10}$ | 3.6x10$^{-15}$ |
| GO | 5.5x10$^4$ | 4.0x10$^{-8}$ | 8.0x10$^9$ | 4.2x10$^{-15}$ |

In summary, we have performed degenerate pump-probe and Z-scan experiments using ~80 fs laser pulses centered at 790 nm on graphene dispersed in three different solvents. Our results clearly demonstrate saturable absorption property and two component carrier relaxation dynamics, with the fast time constant in the range of 130-330 fs and the slower one in 3.5-4.9 ps range. G-H$_2$O which consists primarily of single layered graphene suspended in water shows the largest value of $\beta$ ~9x10$^{-8}$ cm/W. The fastest decay time constant ~130 fs is observed for G-DMF. A quantitative understanding of these differences among different suspension has yet to emerge. The present results demonstrate the fast saturable absorption property of graphene suspensions with possible applications as a passive mode locker in ultrafast lasers and ultrafast optical switches.

A.K.S. thanks Department of Science and Technology, India for financial support and S.K. thanks University Grants Commission, India for Senior Research Fellowship.

[a)]Electronic mail: asood@physics.iisc.ernet.in